\documentclass[prd,preprintnumbers,floatfix,
nofootinbib,superscriptaddress]{revtex4}

\usepackage{float}
\usepackage{nicefrac}
\usepackage{mathtools}
\usepackage{amsfonts} 
\usepackage{amssymb} 
\usepackage{amsmath} 
\usepackage{graphicx} 
\usepackage{subfigure} 
\usepackage{array} 
\usepackage{dcolumn} 
\usepackage{bm} 
\usepackage{latexsym} 
\usepackage{longtable} 
\usepackage{hyperref} 
\usepackage{verbatim}
\usepackage{epsfig}
\usepackage{slashed}
\usepackage{color}

\usepackage{glossaries-extra}
\setabbreviationstyle[acronym]{long-short}
\glssetcategoryattribute{acronym}{nohyperfirst}{true}
\renewcommand{\vec}[1]{\mathbf{#1}}
\newcommand{\cA}[0]{\mathcal A}

\newcommand{\cD}[0]{\mathcal D}

\newcommand{\cK}[0]{\mathcal K}
\newcommand{\cL}[0]{\mathcal L}
\newcommand{\cM}[0]{\mathcal M}
\newcommand{\cO}[0]{\mathcal O}

\newcommand{\cR}[0]{\mathcal R}

\newcommand{\cY}[0]{\mathcal Y}

\newcommand{\Eqref}[1]{(\ref{#1})}

\newcommand{\wt}[0]{\widetilde}

\newcommand{\df}[0]{\mathrm{df}}

\newcommand{\uu}[0]{{(u,u)}}

\newcommand{\Kdf}[0]{{\cK_{\df,3}}}
\newcommand{\Kdfuu}[0]{{\widetilde\cK_{\df,3}^{\uu}}}
\newcommand{\Kdfuun}[0]{{\cK_{\df,3}^{\prime\uu}}}
\newcommand{\KdfuuHS}[0]{{\cK_{\df,3}^{\uu}}}
\newcommand{\PV}[0]{{\mathrm{PV}}}

\newcommand{\MdfuuR}[0]{\cM_{\df,3}^{\cR,\uu}}
\newcommand{\LtoK}[0]{Hansen:2014eka}
\newcommand{\KtoM}[0]{Hansen:2015zga}

\newcommand{\HSTH}[0]{Hansen:2016fzj}
\newcommand{\BHSQC}[0]{Briceno:2017tce}
\newcommand{\BHSnum}[0]{Briceno:2018mlh}
\newcommand{\BHSK}[0]{Briceno:2018aml}
\newcommand{\HSQCa}[0]{Hansen:2014eka}
\newcommand{\HSQCb}[0]{Hansen:2015zga}
\newcommand{\dwave}[0]{Blanton:2019igq}
\newcommand{\largera}[0]{Romero-Lopez:2019qrt}
\newcommand{\HSrev}[0]{Hansen:2019nir}
\newcommand{\BHSSU}[0]{Briceno:2019muc}
\newcommand{\RtoK}[0]{Jackura:2019bmu}
\newcommand{\HHanal}[0]{Blanton:2019vdk}
\newcommand{\isospin}[0]{Hansen:2020zhy}
\newcommand{\BSQC}[0]{BSQC}

\newcommand{\Akakia}[0]{Hammer:2017uqm}
\newcommand{\Akakib}[0]{Hammer:2017kms}

\newcommand{\MDpi}[0]{Mai:2018djl}
\newcommand{\HH}[0]{Horz:2019rrn}

\newcommand{\MD}[0]{Mai:2017bge}

\newcommand{\Akakinum}[0]{Doring:2018xxx}
\newcommand{\Maiisobar}[0]{Mai:2017vot}
\newcommand{\isobar}[0]{Jackura:2018xnx}
\newacronym{CMF}{CMF}{center-of-momentum frame}

\begin{document}


\title{Equivalence of relativistic three-particle quantization conditions}

\author{Tyler D. Blanton}
\email[e-mail: ]{blanton1@uw.edu}
\affiliation{Physics Department, University of Washington, Seattle, WA 98195-1560, USA}

\author{Stephen R. Sharpe}
\email[e-mail: ]{srsharpe@uw.edu}
\affiliation{Physics Department, University of Washington, Seattle, WA 98195-1560, USA}


\date{\today}

\begin{abstract}
We show that a recently derived alternative form of the relativistic three-particle quantization 
condition for identical particles
 can be rewritten in terms of the R matrix introduced to give a unitary representation of the 
infinite-volume three-particle scattering amplitude. Combined with earlier work, this shows
the equivalence 
of the relativistic effective field theory approach of Refs.~\cite{\LtoK,\KtoM}
and the ``finite-volume unitarity'' approach of Refs.~\cite{\MD,\MDpi}.
It also provides a generalization of the latter approach to arbitrary angular momenta of
two-particle subsystems.

 \end{abstract}


\nopagebreak

\maketitle


\section{Introduction\label{sec:intro}}

The study of resonant three-particle systems using lattice QCD (LQCD) is becoming feasible, due to
advances in the underlying theoretical formalism~\cite{Polejaeva:2012ut,\HSQCa,\HSQCb,\BHSQC,\Akakia,\Akakib,\MD,\MDpi,\BHSK,Pang:2019dfe,\largera,\isospin,\BSQC}
and its practical application~\cite{\MDpi,\Akakinum,\BHSnum,\dwave},
as well as in algorithmic and computational methods
necessary to extract three-particle spectra
(see, for example, the recent results presented in Refs.~\cite{\HH,Culver:2019vvu,Beane:2020ycc}).\footnote{%
For related applications to lattice $\phi^4$ theories see Ref.~\cite{Romero-Lopez:2018rcb}.
For alternative approaches see Refs.~\cite{Briceno:2012rv,Guo:2017ism,Klos:2018sen,Guo:2018ibd}.
}
The present frontier is the application to the $3\pi^+$ system~\cite{\HHanal,Mai:2019fba,Culver:2019vvu}.
For recent reviews, see Refs.~\cite{\HSrev,Rusetsky:2019gyk}.

One of the key steps in the formalism is the derivation of three-particle quantization conditions,
equations whose solutions give the finite-volume spectrum of three-particle states as functions 
of infinite-volume two- and three-particle K matrices.
These K matrices can then be related to two- and three-particle scattering amplitudes by solving integral equations.
Three different approaches have been followed to obtain the quantization conditions.

The first is based on an all-orders diagrammatic analysis in a generic relativistic field theory, and is usually
denoted the RFT approach. It was initially developed for identical scalar particles with a G-parity-like $\mathbb Z_2$
symmetry~\cite{\HSQCa,\HSQCb}, and subsequently extended to allow $2\to 3$ processes~\cite{\BHSQC},
the inclusion of poles in the two-particle K matrix~\cite{\BHSK,\largera}, and nonidentical but degenerate
scalars~\cite{\isospin}. In all cases, the formalism allows arbitrary interactions in two-particle subsystems
(which we henceforth refer to as ``dimers''). In a companion paper~\cite{\BSQC}, henceforth referred to as BS1,
we have presented an alternative, simpler, derivation of the RFT quantization condition in the presence of the
$\mathbb Z_2$ symmetry, including an alternative form of the quantization condition itself.
This new form, which depends on an unsymmetrized three-particle K matrix, 
will play a crucial role in the present work. 

The second approach uses nonrelativistic effective field theory (NREFT), allowing a much simplified
derivation of the quantization condition~\cite{\Akakia,\Akakib}. The formalism has so far only been
developed for identical scalars with s-wave dimers and no $2\to3$ transitions.

The third approach, developed in Refs.~\cite{\MD,\MDpi}, 
is based on a unitary parametrization of the three-particle scattering amplitude, $\cM_3$,
in terms of a K-matrix-like real quantity called the R matrix (and denoted $\cR^\uu$ below)~\cite{\Maiisobar,\isobar}.
Following Ref.~\cite{\HSrev}, we call this method the ``finite-volume unitarity'' (FVU) approach.
It leads to a quantization condition that incorporates relativistic effects, and has so far only been developed
for scalars with s-wave dimers and no $2\to3$ transitions.

A natural question is whether there are relations between the approaches, particularly between the
two relativistic approaches (RFT and FVU). 
In addition, as stressed in Ref.~\cite{\HSrev}, it is not clear in the FVU approach whether all sources
of power-law volume dependence have been accounted for.
Thus an alternative derivation of the FVU result would be welcome.

The relationship between approaches was first addressed in Ref.~\cite{\Akakib},
where it was shown that the nonrelativistic limit of the RFT quantization condition of Ref.~\cite{\HSQCa},
restricted to s-wave dimers, reproduced the NREFT result, aside from certain technical differences.
The agreement also required that the quantities describing three-particle interactions in the two approaches
were restricted to their simplest, momentum-independent form.
This agreement was reproduced in Ref.~\cite{\HSrev} using a simplified method.
In addition, Ref.~\cite{\HSrev} showed that, when restricted to s-wave dimers, 
and assuming a constant three-particle interaction, the RFT quantization
condition could be manipulated into a form that agreed with that from the FVU approach
(again aside from certain technical differences).

Our aim here is to extend these results to general two- and three-particle interactions.
In particular, we are able to derive the FVU form of the quantization condition
starting from the RFT result, and thus to generalize the FVU approach to dimers in all partial waves.
The key inputs here are, first, the new form of the RFT quantization condition that we obtained in BS1,
and, second, a generalization we derive here 
of the relation between the K matrix of the RFT approach and the R matrix
obtained  in Ref.~\cite{\RtoK}.
Our final result, given in Eq.~(\ref{eq:QCfinal}), is a form of the quantization condition given explicitly
in terms of $\cR^\uu$.

This article is organized as follows. In the following section we summarize the relativistic
quantization conditions obtained previously,
in both the RFT approach (Sec.~\ref{sec:RFT})
and the FVU approach (Sec.~\ref{sec:FVU}).
Additionally, in Sec.~\ref{sec:RFT} we rewrite the new form of the quantization condition
from BS1 in an alternate form.
In Sec.~\ref{sec:RtoK} we derive the infinite-volume relationship between
asymmetric forms of the three-particle K matrix and the R matrix, $\cR^\uu$.
Using these, in Sec.~\ref{sec:newQC} we rewrite the RFT quantization condition (in its asymmetric form)
in terms of $\cR^\uu$, thus obtaining the general form of the FVU quantization condition.
In a concluding section, Sec.~\ref{sec:outlook},
we briefly compare the advantages of the different  forms of the quantization condition for practical applications.
Appendix~\ref{app:not} summarizes notation and definitions,
while Appendix~\ref{app:inverses} discusses subtleties concerning infinite-volume limits.

\section{Recap of prior forms of the relativistic quantization condition}
\label{sec:QCrecap}

\subsection{Results in the RFT approach}
\label{sec:RFT}

The RFT quantization condition of Ref.~\cite{\HSQCa} is given by
\begin{equation}
\det \left[1 + \Kdf  F_3 \right] = 0\,,
\label{eq:QCHS}
\end{equation}
where $\Kdf$ and $F_3$ are matrices in the space of on-shell three-particle states,
with $F_3$ containing the two-particle K matrix as well as known kinematical factors,
\begin{equation}
F_3 = \wt F \left[ \frac13 - \frac1{\wt H} \, \wt F \right]\,,\qquad
\wt H = 1/\overline{\cK}_{2,L} + \wt F + \wt G\,,
\label{eq:F3}
\end{equation}
while $\Kdf$ is a  three-particle K matrix.
The notation here is that of BS1,
which differs somewhat from that of Ref.~\cite{\HSQCa}.
We summarize the relevant definitions in Appendix~\ref{app:not}, and only note here that
$\overline{\cK}_{2,L}$ 
contains the two-particle K matrix,
while $\wt F$ and $\wt G$ are known kinematic functions.
All three quantities depend on the box size $L$, with the dependence of $\overline{\cK}_{2,L}$
being of a simple kinematic nature [see Eq.~(\ref{eq:K2Lonon})]
while $\wt F$ and $\wt G$ contain the nontrivial volume dependence.
A key property of $\Kdf$ is that it is symmetric under particle exchange, separately 
for both the initial and final three-particle states.
Thus it has the same symmetry properties as the three-particle scattering
amplitude $\cM_3$.

In BS1 we show that the quantization condition of Eq.~(\ref{eq:QCHS}) 
is equivalent to a form written in terms of the asymmetric K matrix $\KdfuuHS$.
Here the right (left) superscript ``$u$'' indicates that one of the three
incoming (outgoing) momenta is being singled out
as being the ``spectator'' in cases where the initial interaction involves only two particles.
The precise definition of $\KdfuuHS$ is given constructively in Ref.~\cite{\HSQCa}, but is not important here.
In fact, to write the asymmetrized quantization condition in a simple form, 
one must use  a new version of the asymmetric K matrix, denoted $\Kdfuun$,
which is obtained from $\KdfuuHS$ by solving an integral equation containing $\cK_2$ and given explicitly
in BS1. Then the new form of the RFT quantization condition is
\begin{equation}
	\det\left[1 + \left(\overline{\cK}_{2,L} + \Kdfuun\right) (\wt{F}+\wt{G}) \right] = 0 \,.
	 \label{eq:QCBS}
\end{equation}
We stress that no information is lost in the transition from $\KdfuuHS$ to $\Kdfuun$, since we do not
have an explicit form for either. In practical applications of the quantization condition, both
must be parametrized. 
They are both related to $\cM_3$ by (different) integral equations,
and both are Lorentz invariant if the relativistic form of $\wt G$ is used.

It turns out to be useful to rewrite the asymmetrized quantization condition as follows:\footnote{
To obtain this form, we are assuming that $\det[\overline\cK_{2,L} + \Kdfuun]\ne 0$, which we expect
to be true in general.
}
\begin{align}
\det\left[\wt H -X^\uu\right] &= 0\,,
\label{eq:QCBSa}
\\
X^\uu &= \overline{\cK}_{2,L}^{-1} -\left[\overline{\cK}_{2,L} + \Kdfuun\right]^{-1}  
\\
&= \overline{\cK}_{2,L}^{-1} \Kdfuun \overline{\cK}_{2,L}^{-1} \frac1{1+\Kdfuun \overline{\cK}_{2,L}^{-1}}\,.
\label{eq:Xuu}
\end{align}
We return below to the issue of whether $X^\uu$ is an infinite-volume object, i.e.~whether
the matrix products in its definition can be replaced by integrals.

BS1 also presents an alternative {\em ab initio} derivation of the asymmetric form of the quantization 
condition,
\begin{equation}
	\det\left[1 + \left(\overline{\cK}_{2,L} + \Kdfuu\right) (\wt{F}+\wt{G}) \right] = 0 \,.
	 \label{eq:QCBSn}
\end{equation}
This differs from Eq.~(\ref{eq:QCBS}) only in the three-particle K matrix that enters: here it is $\Kdfuu$,
while $\Kdfuun$ appears in Eq.~\Eqref{eq:QCBS}.\footnote{%
Another technical difference is that the BS1 derivation defines $\wt G$ using a different boost to the dimer
center-of-mass frame than that used in Ref.~\cite{\HSQCa}. 
However, the derivation of Eq.~(\ref{eq:QCBS}) goes through using either boost.
The equations in the remainder of the paper also hold using either boost---although one must
use the same choice throughout.
}
These two asymmetric K matrices are similar, but differ in their detailed definitions.
$\Kdfuu$ is defined using an asymmetry based on diagrams in time-ordered perturbation theory,
while that for $\Kdfuun$ is based on Feynman perturbation theory, together with additional complications.\footnote{%
One implication of this difference is that $\Kdfuu$ is not Lorentz invariant (irrespective of the choice of
$\wt G$), because it is defined in a frame-dependent way in terms of the diagrams of time-ordered
perturbation theory.
}
 As discussed in BS1, the fact that the same form of the quantization condition can hold with {\em different}
asymmetric K matrices is a reflection of an intrinsic ambiguity in the definition of asymmetric quantities.
We return to this point below.
Finally, we note that Eq.~(\ref{eq:QCBSn}) can also be manipulated into the form of Eq.~(\ref{eq:QCBSa}), 
with $X^\uu$ now given by Eq.~(\ref{eq:Xuu}) with $\Kdfuun$ replaced by $\Kdfuu$.

\subsection{The FVU quantization condition}
\label{sec:FVU}

The FVU form of the quantization condition has been written explicitly so far only when
the particles in the dimer interact in the s wave. The original forms given in Refs.~\cite{\MD,\MDpi}
are quite complicated, but it is shown in  Ref.~\cite{\HSrev} that the FVU quantization condition
can be rewritten as
\begin{equation}
\det\left[\wt H_s - (2\omega L^3)^{-1} \wt C_s^\uu (2\omega L^3)^{-1} \right] = 0 \,,
\label{eq:QCFVU}
\end{equation}
where
$\omega$ is the on-shell single-particle energy (defined in Appendix~\ref{app:not}),
$\wt H_s$ is the s-wave restriction of $\wt H$ (i.e.~with $\ell$ and $m$ set to zero on both sides),
and $\wt C_s^\uu(\vec k,\vec p)$ is a smooth, real function of the spectator momenta.
Using the definition of $\cR^\uu$ given in Refs.~\cite{\MDpi,\RtoK} together with
results from Ref.~\cite{\HSrev}, one can show that $\wt C_s^\uu =  \cR_s^\uu$, 
with $\cR^\uu_s$ the s-wave restriction of $\wt \cR^\uu$.
We have added the ``$(u,u)$" superscript
(which is absent in the original FVU works and in Ref.~\cite{\HSrev})
in order to emphasize that this is an intrinsically asymmetric object, 
since it parametrizes the smooth part of the dimer-particle contact interaction.

In writing the result (\ref{eq:QCFVU}) in terms of $\wt F$ and $\wt G$, 
we are implicitly assuming that we are using
the smooth cutoff function that is built into the approach of Ref.~\cite{\HSQCa}.
The introduction of this cutoff function is essential in that work (and in the alternative approach of BS1) 
in order to argue that all power-law volume dependence is accounted for. 
By contrast, in the FVU approach, a hard cutoff is introduced by hand.
There is, however, no technical reason not to use the smooth cutoff in the FVU approach, and we
assume henceforth that this has been done.

Aside from this technical issue, Eqs.~\Eqref{eq:QCBSa} and \Eqref{eq:QCFVU} are clearly very similar,
and suggest a relation between the s-wave restriction of $X^\uu$ and $\cR_s^\uu$. 
In the following sections we will make this concrete, using a variant of the relationship between $\KdfuuHS$ and the matrix $\cR$ introduced in Ref.~\cite{\RtoK}.

\section{The $\cR^\uu$ matrix and its relation to $\Kdfuun$ and $\Kdfuu$}
\label{sec:RtoK}

One of the results of the RFT approach is an integral equation relating $\Kdf$ to the physical
three-particle scattering amplitude $\cM_3$~\cite{\HSQCb}. 
This provides a representation of $\cM_3$ in terms of a real function 
that is devoid of s-channel unitary cuts (up to the five-particle threshold) and of on-shell singularities.
An important check on this result was the demonstration, in Ref.~\cite{\BHSSU}, that
it provided a representation of $\cM_3$ that satisfied the constraints of s-channel unitarity.\footnote{%
This demonstration remains valid when $\wt G$ is defined with the boost used in BS1.
}
A similar, but different, parametrization of $\cM_3$, in terms of a real K-matrix-like asymmetric%
\footnote{As with $\wt C_s^\uu$, we have added the superscript $(u,u)$, which is not present in the original works, to emphasize the asymmetry of $\cR^\uu$.
}
amplitude $\cR^\uu$, had previously been suggested in the context
of amplitude analyses of experimental results for resonances that decay to three 
particles~\cite{\Maiisobar,\isobar}. 
This parametrization was developed in order to satisfy s-channel unitarity.
In Ref.~\cite{\RtoK}, it was shown that these two parametrizations are equivalent, and the
relationship between $\Kdf$ and $\cR^\uu$ was derived.

Here we need to extend the analysis of Ref.~\cite{\RtoK} to
relate the asymmetric RFT amplitudes $\Kdfuun$ and $\Kdfuu$ to the FVU amplitude $\cR^\uu$.
This brings to light two technical issues that
were overlooked in Ref.~\cite{\RtoK}, although it turns out that they do not impact the final conclusion of that work.
We will describe these in the course of our discussion.

The desired relationships are determined by equating expressions for asymmetric forms
of the three-particle scattering amplitude. We use two such amplitudes:
$\cM_3^\uu$ defined in Ref.~\cite{\HSQCb} in the context of a Feynman diagram analysis,
and $\wt\cM_3^\uu$ defined in BS1 in an analysis using time-ordered perturbation theory (TOPT). 
We present the results for these quantities
in turn, and then compare them to the corresponding expressions in terms of $\cR^\uu$.

\subsection{Expression for $\cM_3^\uu$}

$\cM_3^\uu$ is defined in Ref.~\cite{\HSQCb} using a skeleton expansion in terms of 
Bethe-Salpeter kernels. The external particles can be directly connected to either two- or three-particle kernels.
The asymmetry arises because two-particle kernels are connected to the external momenta
such that the spectator momentum is always associated with the noninteracting propagator.
The connection to the three-particle kernel does not lead to asymmetry, since this kernel is symmetric.

In Ref.~\cite{\HSQCb}, an expression for $\cM_3^\uu$ is obtained
that depends on both $\KdfuuHS$ and $\Kdf$.
In particular, it does not depend solely on the symmetric form $\Kdf$ alone.
This brings up the first technical issue alluded to above.
In the analysis of Ref.~\cite{\RtoK}, a different expression for $\cM_3^\uu$ is used 
that is given wholly in terms of $\Kdf$
[see Eqs.~(20) and (21) of \cite{\RtoK}, in which $\cM_3^\uu$ is called $\cA$].
This is, in fact,  not the correct expression for $\cM_3^\uu$, 
but rather describes a related  (and implicitly defined) quantity,  
in which a certain subclass of diagrams has has been symmetrized.
This change does not impact the final results of Ref.~\cite{\RtoK} because both the correct and incorrect
expressions for $\cM_3^\uu$ symmetrize to the same quantity, $\cM_3$, 
and this is all that is required for the derivation.

Here we use the correct expression for $\cM_3^\uu$. To determine this, we start from the finite-volume version
of the amplitude,
$\cM_{3,L}^\uu$ (also defined in Ref.~\cite{\HSQCb}), which goes over to $\cM_3^\uu$ in the
appropriate $L\to\infty$ limit.
It was shown in BS1 how to asymmetrize the result for $\cM_{3,L}^\uu$ given in Ref.~\cite{\HSQCb}
so as to write it solely in terms of $\KdfuuHS$.
After further manipulation  this is rewritten in BS1 in terms of $\Kdfuun$,
\begin{align}
\cM_{\df,3,L}^\uu &= \cM_{3,L}^\uu - \cD_{L}^\uu
\\
&= 
\frac{1}{1 +\overline{\cK}_{2,L}(\wt F + \wt G)}
\Kdfuun  \frac1{1 + (\wt F+\wt G)\frac{1}{1 + \overline{\cK}_{2,L}(\wt F + \wt G)} \Kdfuun} 
\frac{1}{1 + (\wt F + \wt G)\overline{\cK}_{2,L}} \,.
\label{eq:M3Lc}
\end{align}
Here we have switched to using the divergence-free form of the three-particle amplitude,
whose difference from the original form is given by the multiple two-particle scattering contribution
\begin{equation}
\cD_L^\uu = - \overline{\cM}_{2,L}\wt G \overline{\cM}_{2,L} \frac1{1 + \wt G \overline{\cM}_{2,L}}\,,
\label{eq:DLuu}
\end{equation}
where $\cM_{2,L}$ is defined in Eq.~(\ref{eq:K2LtoM2L}).

Taking the infinite-volume limit of Eq.~\Eqref{eq:M3Lc}
using the $i\epsilon$ prescription described in Ref.~\cite{\HSQCb}, 
we obtain
\begin{align}
\cM_{\df,3}^\uu
&=
\cL\, \Kdfuun 
\frac1{1+  (\wt\rho_\PV + G^\infty)\, \cL\, \Kdfuun}\, \cL^T
\,,
\label{eq:Mdf3uu}
\\
\cL &= \frac1{1 + \overline{\cK}_2 (\wt\rho_\PV +G^\infty)}\,,
\label{eq:LHS}
\\
\cL^T &= \frac1{1 + (\wt\rho_\PV +G^\infty)\overline{\cK}_2 }\,.
\label{eq:LTHS}
\end{align}
This is written in a highly compact notation, adapted from that of Ref.~\cite{\RtoK},
which we now explain. All quantities depend implicitly on initial
and final on-shell variables, each in the $\{\vec k, \ell,  m\}$ space.
For example, $\Kdfuun$ is given explicitly by
\begin{equation}
\Kdfuun(\vec k, \vec p)_{\ell m; \ell' m'} =
\lim_{L\to\infty} \left[\Kdfuun\right]_{k\ell m;p\ell'm'} \,,
\label{eq:Kdfuuprime}
\end{equation}
with $\cM_{\df,3}^\uu$ defined similarly.
The explicit forms for the other quantities are
\begin{align}
\overline{\cK}_2(\vec k,\vec p)_{\ell m;\ell' m'} &=
\lim_{L\to\infty} \left[\overline\cK_{2,L}\right]_{k\ell m;p\ell'm'}
= \overline\delta(\vec k-\vec p) \delta_{\ell \ell'}\delta_{m m'}\; \cK_2^{(\ell)}(q_{2,k}^{*}) \,,
\label{eq:K2bar}
\\
\wt\rho_\PV(\vec k,\vec p)_{\ell m;\ell' m'} &= 
  \overline\delta(\vec k-\vec p) \delta_{\ell \ell'}\delta_{m m'}\; \wt\rho_\PV^{(\ell)}(q_{2,k}^{*2})\,,
\\
G^\infty(\vec k, \vec p)_{\ell m;\ell' m'} &=
\frac{\cY_{\ell m}(\vec p_k^*)}{q_{2,k}^{*\ell}}
\frac{H(\vec k) H(\vec p)}{b_{pk}^2-m^2+i\epsilon}
\frac{\cY_{\ell'm'}(\vec k_p^*)}{q_{2,p}^{*\ell'}}
\,,
\label{eq:Ginfty}
\end{align}
where
\begin{equation}
\overline\delta(\vec k-\vec p) = 2\omega_k (2\pi)^3 \delta^3(\vec k-\vec p)\,,
\end{equation}
and $\cK_2^{(\ell)}$, $\wt\rho_\PV^{(\ell)}$, and the kinematic variables are defined in Appendix~\ref{app:not}.
The products appearing in Eqs.~\Eqref{eq:Mdf3uu}-\Eqref{eq:LTHS} should be viewed as matrix products
in the on-shell index space. Angular momentum indices are summed as usual, while the 
spectator momenta (which are now continuous variables) are integrated with the 
Lorentz-invariant measure\footnote{%
This differs from the notation of  BS1, where the $1/(2\omega_r)$ factor is not
included in the definition of $\int_{\vec r}$.}
$\int_{\vec r} \equiv \int d^3r/(2 \omega_r [2\pi]^3)$.
Thus
\begin{equation}
\left[X Z\right](\vec k, \vec p)_{\ell m;\ell' m'}
\equiv
\sum_{\ell'' m''} \int_{\vec r}
X(\vec k,\vec r)_{\ell m;\ell'' m''}
Z(\vec r,\vec p)_{\ell'' m'';\ell' m'}
\,,
\end{equation}
where $X,Z\in\{\overline{\cK}_2, \wt\rho_\PV, G^\infty, \Kdfuun\}$.
Finally, the inverses in Eqs.~\Eqref{eq:Mdf3uu}-\Eqref{eq:LTHS}, which are well defined as matrix
inverses for finite $L$, become integral equations in the infinite-volume limit. Thus, for example,
$\cL$ satisfies
\begin{equation}
\cL = 1 - \overline{\cK}_2 (\wt\rho_\PV +G^\infty) \cL\,.
\end{equation}

Further details on how the infinite-volume limit of $\cM_{\df,3,L}^\uu$ leads to
Eq.~(\ref{eq:Mdf3uu}) are provided in Appendix~\ref{app:inverses}.
In addition, we describe there how the inverses of $\overline\cK_2$,
$\Kdfuun$, and related quantities are defined, since these are needed below.

\subsection{Expression for $\wt \cM_3^\uu$}

An alternative version of the asymmetric scattering amplitude is introduced in BS1 and denoted 
$\wt \cM_3^\uu$. Its asymmetry is defined in terms of two- and three-particle irreducible
TOPT amplitudes, which differ from the corresponding Bethe-Salpeter kernels. Thus it differs from
$\cM_3^\uu$, although both symmetrize to the physical scattering amplitude $\cM_3$.

The expression for $\wt \cM_3^\uu$ (given in Appendix E of BS1)
is identical to that for $\cM_3^\uu$, Eq.~(\ref{eq:Mdf3uu}),
except with $\Kdfuun$ replaced by $\Kdfuu$:
\begin{equation}
\wt \cM_{\df,3}^\uu
=
\cL\, \Kdfuu 
\frac1{1+  (\wt\rho_\PV + G^\infty) \,\cL \,\Kdfuu} \cL^T
\,.
\label{eq:Mdf3uuBS}
\end{equation}
Here $\Kdfuu$ is the asymmetric K matrix appearing in Eq.~(\ref{eq:QCBSn}),
the new form of the RFT quantization condition obtained in BS1.

\subsection{Result for asymmetric amplitudes in terms of $\cR^\uu$}

We now recall the expression for the asymmetric scattering amplitude in terms of the 
R matrix~\cite{\Maiisobar,\isobar}.
For reasons that will become clear shortly,
we give the amplitude a different name from those discussed earlier, calling it $\MdfuuR$.
We use the form given in Eqs.~(15)-(19) of Ref.~\cite{\RtoK}, which, 
converted into our notation, becomes\footnote{%
In the original works that introduce this form~\cite{\Maiisobar,\isobar},
a different choice of $G^\infty$ was used than that we use here, Eq.~(\ref{eq:Gt}).
In particular, the cutoff function $H(\vec k)$ was replaced with a hard cutoff, 
and barrier factors were not included.
However, as noted in Ref.~\cite{\RtoK}, the derivation of s-channel unitarity---which is the
essential property of this form---goes through for all choices of 
$G^\infty$ that have the same residues of the on-shell poles, 
which is the case for the choices used here.
}
\begin{align}
\MdfuuR
&=
\wt \cL \,\cR^\uu
\frac1{1 - \wt \cL\, \cR^\uu} \wt\cL
\,,
\label{eq:Mdf3uuR}
\\
\wt \cL
&=  \overline{\cM}_2 \frac1{1 + G^\infty \overline{\cM}_2}
= \frac1{1 + \overline{\cM}_2 G^\infty} \overline{\cM}_2 
\,,
\label{eq:M2bar}
\end{align}
where
\begin{equation}
\overline{\cM}_2(\vec k,\vec p)_{\ell m;\ell' m'} = 
\overline\delta(\vec k-\vec p) \delta_{\ell \ell'}\delta_{m m'}\; \cM_2^{(\ell)}(q_{2,k}^{*}) \,,
\end{equation}
with $\cM_2^{(\ell)}$ being the $\ell$th partial wave of the two-particle scattering amplitude.
Using the result
\begin{equation}
\overline{\cM}_2 = \overline{\cK}_2 \frac1{1+ \wt \rho_\PV \overline{\cK}_2}\,,
\end{equation}
which follows from Eq.~(\ref{eq:K2toM2}), we find
\begin{equation}
\wt \cL=
\frac1{1+\overline\cK_2 (\wt\rho_\PV+G^\infty)} \overline\cK_2
= \overline\cK_2 \frac1{1+(\wt\rho_\PV+G^\infty) \overline\cK_2}\,.
\end{equation}

Before comparing to the earlier expressions \Eqref{eq:Mdf3uu} and \Eqref{eq:Mdf3uuBS}, 
we discuss the second technical issue alluded to above.
This issue is whether $\MdfuuR$ should be equated to
$\cM_{\df,3}^\uu$ or to $\wt \cM_{\df,3}^\uu$.
All three amplitudes symmetrize to the same quantity, $\cM_{\df,3}$,
but this does not guarantee equivalence before symmetrization.
Furthermore, as we have already noted, the analysis of Ref.~\cite{\RtoK} uses a different,
partially symmetrized version of $\cM_{\df,3}^\uu$ (which also symmetrizes to $\cM_{\df,3}$).
In Ref.~\cite{\RtoK}, it is implicitly assumed that this last version of the asymmetric amplitude
is equal to $\MdfuuR$.
However, since the R-matrix parametrization is not obtained using Feynman or TOPT diagrams, 
but rather is a form constructed solely to satisfy s-channel unitarity, 
we see no {\em fundamental} way of connecting it to any of the diagram-based definitions.
We also see no sense in which either $\cM_{\df,3}^\uu$ or $\wt \cM_{\df,3}^\uu$ 
(or the partially symmetrized version of the former) is better suited to an R-matrix parametrization.

We propose that the resolution to this conundrum is that $\cR^\uu$ is intrinsically ambiguous,
and that, with suitable choices of this quantity, we can equate $\MdfuuR$ to either $\cM_{\df,3}^\uu$
or $\wt \cM_{\df,3}^\uu$ (or to the partially symmetrized version of the former, as done in Ref.~\cite{\RtoK}).
To say it differently, we propose that the parametrization of 
$\cM_3$ in terms of $\cR^\uu$ involves a redundancy,
such that a family of choices of $\cR^\uu$ leads to the same physical scattering amplitude.
We stress that we are not suggesting that any ambiguity arises in the relation between $\cR^\uu$
and $\MdfuuR$---for a given choice of the latter quantity (including its subthreshold continuation),
we expect that $\cR^\uu$ is uniquely determined.
The ambiguity arises in the definition of $\MdfuuR$ itself.\footnote{%
A potentially confusing point is that, in Ref.~\cite{\RtoK}, the amplitude $\MdfuuR$ is called
$\cA_{\vec p' \vec p}$, with no explicit indication that it is an asymmetric quantity.
We stress that $\cA_{\vec p' \vec p}$ is asymmetric, 
and is related to $\cM_3$ by the symmetrization procedure of Eq.~(7) of Ref.~\cite{\RtoK}.
}
The relations derived below demonstrate {\em a posteriori} the validity of our proposal,
because they show that the expressions given above for both $\cM_{\df,3}^\uu$
and $\wt \cM_{\df,3}^\uu$ can be rewritten exactly in the R-matrix form.

\subsection{Combining results}
\label{sec:combine}

Returning to the main line of argument, we note that the external integral operators in the two
expressions, Eqs.~\Eqref{eq:Mdf3uu} and \Eqref{eq:Mdf3uuR}, are related by
\begin{equation}
\wt \cL = \cL\, \overline\cK_2 = \overline\cK_2\, \cL^T\,.
\end{equation}
Thus, Eq.~\Eqref{eq:Mdf3uuR} can be rewritten as\footnote{%
The inverses appearing in this section and the next are defined in Appendix~\ref{app:inverses}.
}
\begin{align}
\MdfuuR
&=
 \cL\, \overline\cK_2 \,\cR^\uu
\frac1{1 -  \cL \,\overline\cK_2 \,\cR^\uu} \overline\cK_2 \,\cL^T
\\
&= \cL \frac1{\overline\cK_2^{-1} [\cR^\uu]^{-1} \overline\cK_2^{-1} - \overline\cK_2^{-1} \cL} \cL^T\,.
\end{align}
Comparing this to a slightly rewritten version of Eq.~\Eqref{eq:Mdf3uu}
\begin{align}
\cM_{\df,3}^\uu
&=
\cL \frac1{\big[\Kdfuun\big]^{-1} + (\wt\rho_\PV+G^\infty) \cL} \cL^T\,,
\end{align}
we observe that these expressions match if and only if
\begin{align}
\Big[\Kdfuun\Big]^{-1} &= \overline\cK_2^{-1} \big[\cR^\uu\big]^{-1} \overline\cK_2^{-1} - \Big(\overline\cK_2^{-1} +\wt\rho_\PV+G^\infty\Big) \cL 
\\
&= \overline\cK_2^{-1} \big[\cR^\uu\big]^{-1} \overline\cK_2^{-1} - \overline\cK_2^{-1}
\,,
\label{eq:RtoKdf}
\end{align}
where the second step follows from Eq.~(\ref{eq:LHS}).
This can be rewritten as
\begin{equation}
\Kdfuun = \overline\cK_2 \cR^\uu \overline\cK_2 \frac1{1-\cR^\uu \overline\cK_2}\,,
\label{eq:RtoK}
\end{equation}
or, equivalently, as an integral equation
\begin{equation}
\Kdfuun = \overline\cK_2 \cR^\uu \overline\cK_2 + \overline\cK_2 \cR^\uu \Kdfuun\,.
\end{equation}
The inverse relation can also be given, as discussed below.
Reversing the algebraic steps, we conclude that, if $\Kdfuun$ and $\cR^\uu$ are related in this manner, 
then $\cM_{\df,3}^\uu$ can be written in the R-matrix form of Eq.~\Eqref{eq:Mdf3uuR}.

We can follow exactly the same steps 
if we equate the result for $\wt \cM_{\df,3}^\uu$, Eq.~\Eqref{eq:Mdf3uuBS}, to $\MdfuuR$.
Thus, with a {\em different} choice of $\cR^\uu$, we have
\begin{equation}
\Kdfuu = \overline\cK_2 \cR^\uu \overline\cK_2 \frac1{1-\cR^\uu \overline\cK_2}\,.
\label{eq:RtoKn}
\end{equation}

The relations (\ref{eq:RtoK}) and (\ref{eq:RtoKn}) are
simpler than that between (a third choice of) $\cR^\uu$ and $\Kdf$ obtained in Ref.~\cite{\RtoK}.
This is perhaps to be expected as both are asymmetric quantities.
We note that the new relations are consistent with the fact that both the R and  K matrices are purely real.
The appearance of factors of $\overline\cK_2$ ``wrapping'' $\cR^\uu$ is a result of the choice in the
R-matrix approach of pulling out the dimer scattering amplitude as an explicit external factor---see
Fig.~2(a) of Ref.~\cite{\RtoK}.

A technical point concerns the integrals over intermediate momenta
that are implicit in Eqs.~(\ref{eq:RtoK}) and (\ref{eq:RtoKn}).
Expanding the geometric series, there are terms of the form $\dots \cR^\uu \overline\cK_2 \cR^\uu \dots$,
which lead to an integral over the spectator-momentum associated with $\cK_2^{(\ell)}$.
If there are narrow resonances in a given channel, then $\cK_2^{(\ell)}$ can have poles on the real axis, 
and one must specify how to do the integrals. 
These can be dealt with either by using a pole prescription or by generalizing the
principal value (PV) prescription used to define $\cK_2^{(\ell)}$, which can move the poles
out of the relevant kinematic range~\cite{\largera}.
We prefer the latter approach, as this generalized PV prescription is needed to derive the
quantization condition of Eq.~\Eqref{eq:QCBSa} in the case where $\cK_2^{(\ell)}$ has poles.

In fact, although $\cK_2^{(\ell)}$ and $\Kdfuun$ both depend  on the choice of PV prescription,
it turns out that all choices of $\cR^\uu$ are prescription independent.
The key fact here is that the combination $\overline{\cK}_{2,L}^{-1} + \wt F$ is, by construction,
independent of the prescription. 
This in turn implies that $\wt \cL$ is also prescription independent,
since it can be written
\begin{equation}
\wt \cL =\lim_{L\to\infty} \frac1{ \overline{\cK}_{2,L}^{-1} + \wt F + \wt G}
\,.
\end{equation}
Finally, using Eq.~(\ref{eq:Mdf3uuR}) and the fact that $\MdfuuR$ is prescription independent
(which follows from the prescription independence of $\cM_{3}$ and $\cD^\uu$), 
we see that $\cR^\uu$ must also be independent of the PV prescription.
In this sense, $\cR^\uu$ is a ``more physical'' quantity than $\Kdfuun$ or $\Kdfuu$.
We note, however, that $\cR^\uu$ does depend on the cutoff function, 
since that dependence enters through $G^\infty$ and is not canceled.

\section{Expressing the quantization condition in terms of $\cR^\uu$}
\label{sec:newQC}

We are now ready to combine the results obtained above to rewrite the quantization condition in terms
of $\cR^\uu$. For definiteness, we first consider the choice of $\cR^\uu$ that is related to $\Kdfuun$
by Eq.~(\ref{eq:RtoK}), and thus consider the form of the quantization condition containing the latter quantity,
Eq.~(\ref{eq:QCBSa}). We discuss the other choices of $\cR^\uu$ subsequently.

We start from Eq.~(\ref{eq:RtoKdf}), from which follows
\begin{align}
	\big[\cR^\uu\big]^{-1} &= \overline\cK_2 + \overline\cK_2 \Big[\Kdfuun\Big]^{-1} \overline\cK_2\,.
\end{align}
This can be rewritten as
\begin{align}
\cR^\uu &= \overline{\cK}_2^{-1} - \left[\overline{\cK}_2 + \Kdfuun \right]^{-1} 
\\
&= \overline{\cK}_2^{-1} \Kdfuun \overline{\cK}_2^{-1} \frac1{1+\Kdfuun \overline{\cK}_2^{-1}}\,.
\end{align}
The key observation is that the quantity $X^\uu$ appearing in the quantization condition, Eq.~(\ref{eq:Xuu}),
satisfies
\begin{align}
\lim_{L\to\infty} (2\omega L^3) X^\uu (2\omega L^3) &= 
\cR^\uu\,,
\label{eq:KtoR}
\end{align}
where the factors of $(2\omega L^3)$ arise from Eq.~\Eqref{eq:K2invlim}.
It follows that, if the finite-volume corrections to this result are exponentially suppressed,
i.e.~if
\begin{align}
\left[(2\omega L^3)X^\uu (2\omega L^3)\right]_{k\ell m;p \ell' m'} &= 
\left[\cR^\uu\right]_{k\ell m; p \ell' m'} + \cO(e^{-mL})\,,
\label{eq:needed}
\end{align}
then the quantization condition \Eqref{eq:QCBSa} can be rewritten as
\begin{equation}
\det\left[ \wt H - (2\omega L^3)^{-1} \cR^\uu (2\omega L^3)^{-1}\right] = 0\,.
\label{eq:QCBSFVU}
\end{equation}
Here $\cR^\uu$ is the matrix form of the infinite-volume amplitude, obtained in the usual
way
\begin{equation}
\left[\cR^\uu\right]_{k\ell m; p \ell' m'}  \equiv \cR^\uu(\vec k,\vec p)_{\ell m;\ell' m'}\,, \qquad
\{\vec k,\vec p\} \in (2\pi/L) \mathbb Z^3\,,
\end{equation}
i.e. by restricting the momenta to the finite-volume set.

To discuss the validity of Eq.~(\ref{eq:needed}), we consider the definition of $X^\uu$,
Eq.~(\ref{eq:Xuu}).
Expanding out the geometric series, we find terms of the form 
$\dots\Kdfuun \overline{\cK}_{2,L}^{-1} \Kdfuun\dots$.
As shown in Eq.~(\ref{eq:KKinvKlim}) this goes over to $\dots \Kdfuun \overline{\cK}_2^{-1} \Kdfuun\dots$
in the infinite-volume limit, with the intermediate momentum sums over spectator momenta converted
to integrals.
However, if $\cK_{2}^{(\ell)}$ has zeros within the kinematic range of interest (which ranges up to the four pion
threshold for two-particle scattering), then the difference between sum and integral
over the resulting poles in $\cK_2^{-1}$ will lead to
power-law corrections to Eq.~(\ref{eq:needed}), which would invalidate the quantization condition
(\ref{eq:QCBSFVU}). Zeros in $\cK_2^{(\ell)}$ (along the real $q_{2,k}^{*2}$ axis) occur when the
phase shift passes through $n \pi$ with $n\in\mathbb{Z}$ and have no particular physical significance.
Excluding such cases would be a major restriction on the applicability of Eq.~(\ref{eq:QCBSFVU}).

In fact, we do not think that such cases need to be excluded.
The point is that we expect $\cR^\uu$ to be finite in the vicinity of positions where $\cK_2^{(\ell)}$
(and thus $\cM_2^{(\ell)}$) has zeros. 
This is because, as noted above, $\cR^\uu$ is defined in the expression for $\cM_{\df,3}$ 
with factors of $\cM_2^{(\ell)}$ pulled out on both sides
[as can be seen from Eq.~\eqref{eq:Mdf3uuR}].
Thus the effects of a vanishing $\cM_2^{(\ell)}$ are already included.
Assuming so, then Eq.~(\ref{eq:RtoK}) shows that $\Kdfuun$ vanishes at such positions---specifically,
$\Kdfuun(\vec k,\vec p)_{\ell m;\ell' m'} = 0$ if $\cK_2(\vec k)_{\ell m}=0$ or $\cK_2(\vec p)_{\ell' m'}=0$.
This implies that the divergences in $\overline{\cK}_2^{-1}$ occurring in the expression
for $X^\uu$ are canceled by the behavior of $\Kdfuun$.
Thus we conclude that Eq.~\Eqref{eq:QCBSFVU} is a legitimate form of the quantization condition.

We can repeat the arguments just given using the quantization condition written in terms of
$\Kdfuu$, Eq.~(\ref{eq:QCBSn}), and the relation between $\Kdfuu$ and a different choice for $\cR^\uu$
given in Eq.~(\ref{eq:RtoKn}). The result is that the quantization condition can be written
in exactly the form of Eq.~\Eqref{eq:QCBSFVU}, except with the new choice of $\cR^\uu$.
One disadvantage of this choice of $\cR^\uu$ is that it is not Lorentz invariant. This follows because
it is defined in terms of the TOPT asymmetric amplitude $\wt \cM_3^\uu$, which depends on the
choice of frame used to define the time axis.
By contrast, the form of $\cR^\uu$ obtained by equating $\cM_{\df,3}^{\cR,\uu}$
to $\cM_{\df,3}^\uu$ is Lorentz invariant,
as long as the relativistic form of $\wt G$ is used.

\section{Summary and outlook}
\label{sec:outlook}

The main result of this work is the demonstration that the three-particle quantization condition
for scalar particles with a $\mathbb Z_2$ symmetry
obtained in the RFT approach in Ref.~\cite{\HSQCa} (and extended in BS1)
can be rewritten in terms of the R matrix of Refs.~\cite{\Maiisobar,\isobar}
in the simple form
\begin{equation}
\det\left[ \overline{\cK}_{2,L}^{-1}+ \wt F + \wt G  - (2\omega L^3)^{-1} \cR^\uu (2\omega L^3)^{-1} \right] = 0\,.
\label{eq:QCfinal}
\end{equation}
This provides the generalization of the s-wave FVU result of Refs.~\cite{\MD,\MDpi},
Eq.~(\ref{eq:QCFVU}), to all angular momenta of the dimer,
and shows the equivalence of the RFT and FVU approaches in general.\footnote{%
As noted earlier, Eq.~(\ref{eq:QCFVU}) is obtained from the original result for the FVU quantization condition,
given in Refs.~\cite{\MD,\MDpi}, only after some algebraic manipulations~\cite{\HSrev}.
Presumably, our generalized result  could be rewritten in a form
similar to that of the original works, but we have not attempted this.
}
We stress that the derivation of Eq.~(\ref{eq:QCfinal}) requires the use of a smooth cutoff function
(as opposed to a hard cutoff) as well as
the presence of the ``barrier factors'' in the 
definition of $\wt G$ [see discussion below Eq.~(\ref{eq:Gt})].
We note that, while the two-particle interaction enters with a factor of $1/L^3$ (contained in
$\overline\cK_{2,L}$), the three-particle interaction term comes with a $1/L^6$.
This is as expected based on the overlap amplitudes of particles with wavefunctions distributed
throughout the volume, and is consistent with the results of the
threshold expansion~\cite{Tan:2007bg,Beane:2007qr,\HSTH}.
We expect that by taking the nonrelativistic limit of this form of the quantization condition, one
will obtain the generalization of the NREFT quantization condition of Refs.~\cite{\Akakia,\Akakib}
to all dimer angular momenta.

We have also found that the R matrix is not unique, but rather that Eq.~(\ref{eq:QCfinal}) holds for two different
choices of $\cR^\uu$, which are in turn related to the two different asymmetric forms of the three-particle
K matrix that we have discussed, namely $\Kdfuun$ and $\Kdfuu$. We have argued that the lack of
uniqueness of $\cR^\uu$ is an example of the general result that asymmetric forms of amplitudes are
intrinsically ambiguous, since the process of symmetrization is not invertible. This is most obviously
seen in the fact that one can consider two different asymmetric forms of the three-particle scattering amplitude,
$\cM_3^\uu$ and $\wt \cM_3^\uu$, whose definitions differ by whether the asymmetry is defined with respect
to a Feynman-diagram-based skeleton expansion~\cite{\HSQCb} or an expansion in terms of time-ordered
perturbation theory (see BS1). 

Looking forward, an important question is how the new, asymmetric form of the quantization condition,
Eq.~(\ref{eq:QCfinal}), compares in practice with the original, symmetric form of Eq.~(\ref{eq:QCHS}).
The advantages of the new form include its simplicity and the fact that $\cR^\uu$ is independent of
the choice of PV prescription. It is also closely connected to phenomenological analyses of
scattering amplitudes, through which intuition and experience concerning appropriate parametrizations
of $\cR^\uu$ have been developed.
The disadvantage of the new form is that $\cR^\uu$ is an asymmetric amplitude, whose general description
requires additional parameters in comparison to the symmetric K matrix $\Kdf$ that enters Eq.~(\ref{eq:QCHS}).
This is clear, for example, in the threshold expansion worked out in Ref.~\cite{\dwave}, where a significant
reduction in parameters occurs because of the symmetry of $\Kdf$.\footnote{%
One can also see this in the result for $\KdfuuHS$ one obtains in leading-order chiral perturbation theory,
by extending the calculations described in Ref.~\cite{\HHanal}.}

\section*{Acknowledgments}

We thank Max Hansen, Andrew Jackura and Maxim Mai for comments on the manuscript and discussions.
This work was supported in part by the U.S. Department of Energy contract DE-SC0011637.

\appendix
\section{Summary of notation and definitions}
\label{app:not}

We collect here the definitions of quantities used in the main text, and explain the notation that is used.
These results are drawn from Refs.~\cite{\HSQCa,\HSQCb} and \cite{\BSQC}, and we use the notation of
the latter work (referred to as BS1 in the main text). We present a bare-bones description here---see
these references for further details.

Configurations of three on-shell particles are described by denoting one of the particles
as the spectator and the other two as the interacting pair, or ``dimer'' for short.
These designations are intrinsically asymmetric, and this is used when defining the asymmetric kernels
such as $\cR^\uu$ and $\Kdfuun$, where any initial two-particle interaction is always chosen to involve the dimer, 
although subsequently all three particles interact.
For symmetric quantities such as $\Kdf$, the choice of which momentum is denoted the spectator
is irrelevant.

In more detail, for a given total 4-momentum $P^\mu=(E,\vec P)$, 
the momentum dependence of quantities is specified by
giving the momentum of the spectator, call it $\vec k$, 
and then boosting to the center-of-mass frame (CMF) of the dimer and decomposing
the momentum dependence of one particle in the dimer into spherical harmonics.
Thus the variables are $\{\vec k,\ell, m\}$. A similar set of variables is used for both initial and final
momenta, so that, for example, the on-shell scattering amplitude can be written
$\cM_3(\vec k,\vec p)_{\ell m;\ell' m'}$. In infinite volume, $\vec k$ and $\vec p$ are continuous variables.

In the quantization conditions, the spectator momenta are constrained by the boundary conditions,
here chosen to be periodic in the box size $L$. Then $\vec k=\frac{2\pi}{L}\vec n$, with $\vec n\in\mathbb{Z}^3$.
Thus all of the variables become discrete, and we denote the full set by $\{k\ell m\}$, with $k$ a shorthand
for the discrete choices of $\vec k$. All quantities in the quantization condition are then matrices in which
each of the indices runs over the set $\{k\ell m\}$. For quantities that are initially defined in infinite volume,
the restriction to the finite-volume matrix versions is exemplified by
\begin{equation}
\cR^\uu_{k\ell m;p \ell' m'} = \cR^\uu(\vec k,\vec p)_{\ell m;\ell' m'}\,,\qquad
\{\vec k, \vec p\} \in \frac{2\pi}{L} \mathbb Z^3 \,.
\end{equation}

Two-particle quantities that enter the quantization condition naturally come with associated factors
of $2\omega_k L^3$, where $\omega_k=\sqrt{\vec k^2+m^2}$, with $m$ the particle mass. 
These quantities are overlined,
to distinguish them from the infinite-volume two-particle amplitude that they contain, and given
a subscript ``$L$'' to emphasize their volume dependence. For example,
\begin{align}
\left[\overline{\cK}_{2,L}\right]_{k\ell m;p\ell' m'} &=  \left[(2\omega L^3)\cK_2\right]_{k\ell m;p\ell' m'}\,,
\label{eq:K2Lonon}
\\
\left[\cK_2\right]_{k\ell m;p\ell' m'} &= \delta_{kp}\delta_{\ell \ell'}\delta_{m m'} \cK_2^{(\ell)}(q_{2,k}^*)\,,
\label{eq:K2on}
\\
\left[2\omega L^3\right]_{k\ell m;p \ell' m'} &= \delta_{kp} \delta_{\ell \ell'} \delta_{m m'} 2 \omega_k L^3\,,
\label{eq:2omL3}
\end{align}
where $\cK_2^{(\ell)}$ is the $\ell$th partial wave of the infinite-volume two-particle K matrix, which depends
on the dimer CMF relative momentum, 
\begin{equation}
q_{2,k}^{*} = \sqrt{E_{2,k}^{*2}/4-m^2} \,,
\qquad
E_{2,k}^{*2} = (E-\omega_k)^2 - (\vec P-\vec k)^2\,.
\end{equation}
Following Refs.~\cite{\HSQCa,\largera}, we define $\cK_2$ using a generalized principal value (PV) 
pole prescription, such that its relation to the physical two-particle scattering amplitude $\cM_2$ is
\begin{equation}
 \left[\cK_2^{(\ell)}(q_{2,k}^*)\right]^{-1}
=
\left[\cM_2^{(\ell)}(q_{2,k}^*)\right]^{-1}
-
 \wt\rho_\PV^{(\ell)}(q_{2,k}^{*2})
\,,
\label{eq:K2toM2}
\end{equation}
where 
\begin{equation}
\wt\rho_\PV^{(\ell)}(q_{2,k}^{*2}) = 
H(\vec k) \left[\wt\rho(q_{2,k}^{*2}) + \frac1{32\pi^2}I_{\PV}^{(\ell)}(q_{2,k}^{*2}) \right]\,,
\label{eq:rhoPVdef}
\end{equation}
with the phase space factor given by
\begin{equation}
\wt\rho(q_{2,k}^{*2}) = \frac1{16\pi E_{2,k}^*} \left\{
\begin{array}{lr} -i {|q_{2,k}^{*}|} &\quad q_{2,k}^{*2} > 0 \\
\hphantom{-i}| q_{2,k}^{*}| &\quad q_{2,k}^{*2} \le 0 \end{array}\right.\,.
\label{eq:rhodef}
\end{equation}
$I_{\PV}^{(\ell)}$ is an arbitrary real, smooth function, which is used to move poles in $\cK_2$
out of the kinematic range of interest.
$H(\vec k)$ is a smooth cutoff function, which cuts off the sum over $\vec k$ for $|\vec k|\sim m$.
Examples are given in Refs.~\cite{\HSQCa,\BHSQC}. 

The kinematic functions $\wt F$ and $\wt G$ are
\begin{align}
	 \wt{F}_{k\ell m;p\ell'm'}  &=  \delta_{kp}\frac{H(\vec k)}{2\omega_k L^3}
	\left[ \frac1{L^3} \sum_{\vec{a}}^{\rm UV} - \PV\!\int^{\rm UV}\!\! \!\! \frac{d^3a}{(2\pi)^3}\right] 
	\frac{\cY_{\ell m}(\vec{a}_{k}^*)}{q_{2,k}^{*\ell}} 
	\frac1{2! 2 \omega_a (b_{ka}^2-m^2)}
	\frac{\cY_{\ell'm'}(\vec{a}_{k}^{*})}{q_{2,k}^{*\ell'}}
	\,,  \label{eq:Ft}
\\
\wt G_{k\ell m;p\ell' m'}
&=
\frac1{2\omega_kL^3} \frac{\cY_{\ell m}(\vec{p}_k^*)}{q_{2,k}^{*\ell}}
\frac{H(\vec k)H(\vec p)}{b_{pk}^2-m^2}
\frac{\cY_{\ell'm'}(\vec{k}_p^*)}{q_{2,p}^{*\ell'}} \frac1{2\omega_p L^3}\,.
\label{eq:Gt}
\end{align}
Here $b_{ka}^\mu \equiv P^\mu -k^\mu-a^\mu$ is a four vector, with $k^\mu=(\omega_k,\vec k)$
and $a^\mu=(\omega_a,\vec a)$, and $b_{kp}$ is defined analogously.
The sum over $\vec a$ in $\wt F$ runs over the finite-volume set.
Momenta with an asterisk, e.g.~$\vec a_k^*$ and $\vec p_k^*$, are boosted from the original frame
(with total momentum $\vec P$) into the CMF of the dimer.
There is some flexibility in the choice of boost, with two examples being given in Refs.~\cite{\HSQCa} and BS1.
The former leads to a relativistically covariant $\wt G$, in the sense that the vectors
$\vec p_k^*$ and $\vec k_p^*$ are unchanged (up to global rotations) when the initial frame is 
given an arbitrary boost.
We refer to this form of $\wt G$ as the ``relativistic form'' in the main text.
The harmonic polynomials are defined by
\begin{equation}
\cY_{\ell m}(\vec{a})
= \sqrt{4\pi} Y_{\ell m}(\widehat{\vec a}) |\vec a|^\ell\,,
\end{equation}
with the spherical harmonics chosen to be in the real basis.
The factors of $|\vec p_k^*|^\ell/q_{2,k}^{*\ell}$ in $\wt G$ are called ``barrier factors'' in the main text,
and are needed to assure the smoothness of $\wt G$ when $|\vec p_k^*|\to 0$.
The superscript UV on the sum and integral in $\wt F$ indicate an ultraviolet regularization, the nature
of which affects $\wt F$ only at the level of exponentially suppressed terms.
Finally, the integral in $\wt F$ is defined by the generalized pole prescription mentioned above~\cite{\largera}.

With these definitions, the finite-volume two-particle scattering amplitude (defined in Ref.~\cite{\HSQCb})
is given by
\begin{equation}
\left[\overline{\cM}_{2,L}\right]^{-1}
=
 \left[\overline{\cK}_{2,L}\right]^{-1}
+
 \wt F
\,,
\label{eq:K2LtoM2L}
\end{equation}
which in the appropriate $L\to\infty$ limit goes over to Eq.~(\ref{eq:K2toM2}).

\section{Infinite-volume limits}
\label{app:inverses}

In this appendix we provide further details of the infinite-volume limit needed to obtain Eq.~\Eqref{eq:Mdf3uu}
from Eq.~\Eqref{eq:M3Lc}, and discuss the properties of the inverses that appear in
Secs.~\ref{sec:combine} and \ref{sec:newQC}.

To obtain Eq.~\Eqref{eq:Mdf3uu} we need to show that
\begin{align}
	\lim_{L\to\infty} X_L (\wt F + \wt G) Z_L &= 
	\lim_{L\to\infty} X_L \frac1{(2\omega L^3)} (2\omega L^3) (\wt F + \wt G) 
	(2 \omega L^3) \frac1{(2\omega L^3)} Z_L 
\\
&=	X_\infty (\wt\rho_\PV + G^\infty) Z_\infty\,,
\label{eq:rhs}
\end{align}
with $X_L,Z_L\in\{\Kdfuun,\overline\cK_{2,L}\}$ being finite-volume matrices,
and $X_\infty,Z_\infty$ their corresponding infinite volume limits,
given in Eqs.~(\ref{eq:Kdfuuprime}) and (\ref{eq:K2bar}).
The factors of $1/(2\omega L^3)$ convert the sums over intermediate momenta into the Lorentz-invariant
integrals that are implicit in the infinite-volume form, Eq.~(\ref{eq:rhs}).
The remainder of the result can be obtained using
\begin{align}
\lim_{L\to\infty} 2\omega_k L^3\wt F_{k\ell m;p\ell'm'} 2\omega_p L^3
&= \wt\rho_\PV(\vec k,\vec p)_{\ell m;\ell'm'} 
	\\
\lim_{L\to\infty} 2 \omega_k L^3 \wt G_{k\ell m;p\ell'm'} 2\omega_p L^3
&=
G^\infty(\vec k,\vec p)_{\ell m;\ell'm'} \,.
\end{align}
The first line follows from the results in Appendix B of BS1, while the second follows from
the definitions of $\wt G$, Eq.~(\ref{eq:Gt}), and $G^\infty$, Eq.~(\ref{eq:Ginfty}).

We now turn to the definition of inverses of infinite-volume quantities, beginning with
$\overline{\cK}_2^{-1}$.
Given our integration measure, this should satisfy
\begin{equation}
\sum_{\ell'' m''}\int_{\vec r} \overline{\cK}_2^{\;-1}(\vec k,\vec r)_{\ell m;\ell'' m''}
 \overline{\cK}_2(\vec r, \vec p)_{\ell' m'';\ell' m'} = \overline\delta(\vec k-\vec p)\delta_{\ell \ell'} \delta_{m m'}\,,
 \label{eq:K2invinfty}
 \end{equation}
 from which it follows that
\begin{align}
	\overline{\cK}_2^{-1}(\vec k,\vec p)_{\ell m;\ell' m'} &\equiv 
\overline \delta(\vec k-\vec p)\delta_{\ell \ell'}\delta_{m m'}\; \left[\cK_2^{(\ell)}(q_{2,k}^{*})\right]^{-1} \,.
\end{align}
A drawback of our notation is that, although the infinite-volume limit of $\overline{\cK}_{2,L}$,
given in Eq.~(\ref{eq:K2bar}), looks natural,
the same is not true of the inverse
\begin{equation}
\lim_{L\to\infty} (2\omega L^3) \left[\overline{\cK}_{2,L} \right]^{-1} (2\omega L^3)  = \overline{\cK}_2^{-1}\,.
\label{eq:K2invlim}
\end{equation}
The extra factors of $(2\omega L^3)$ are, however, needed so that the
limit of matrix products is as expected. For example, we have
\begin{equation}
\lim_{L\to \infty}\left[\Kdfuun \overline{\cK}_{2,L}^{-1} \Kdfuun\right]_{k\ell m;p\ell' m'}
=
\left[\Kdfuun \overline{\cK}_2^{-1} \Kdfuun\right](\vec k,\vec p)_{\ell m;\ell' m'}
\,,
\label{eq:KKinvKlim}
\end{equation}
with the matrix multiplications converted to Lorentz-invariant integrals by the 
induced factors of $(2\omega L^3)^{-1}$.
The only places where these extra factors are not absorbed are on the ends of expressions,
such as in Eq.~(\ref{eq:KtoR}).

The inverses of $\Kdfuun$ and $\cR^\uu$ appearing in Sec.~\ref{sec:newQC}
are defined as in Eq.~(\ref{eq:K2invinfty}), e.g.
\begin{equation}
\sum_{\ell'' m''}\int_{\vec r} [\cR^\uu]^{-1}(\vec k,\vec r)_{\ell m;\ell'' m''}
\cR^\uu(\vec r, \vec p)_{\ell' m'';\ell' m'} = \overline\delta(\vec k-\vec p)\delta_{\ell \ell'} \delta_{m m'}\,,
 \end{equation}
The relation of these inverses to the inverses of their finite-volume versions display
similar peculiarities to that seen in Eq.~(\ref{eq:K2invlim}), but these relations are not needed
in the arguments of the main text.

\bibliography{ref} 

\end{document}